\documentclass[pra,twocolumn,nofootinbib,eqsecnum,floatfix,showpacs]{revtex4}
\usepackage{bm} 
\usepackage{amsmath}

\begin{document}

\title{{\bf Nonvanishing Local Scalar Invariants even in VSI Spacetimes\\
with all Polynomial Curvature Scalar Invariants Vanishing}
\thanks{Alberta-Thy-15-08, arXiv:0806.2144}}

\author{Don N. Page}
\email{don@phys.ualberta.ca}

\affiliation{Theoretical Physics Institute\\
Department of Physics, University of Alberta\\
Room 245B1 CEB, 11322 -- 89 Avenue\\
Edmonton, Alberta, Canada T6G 2G7}

\date{2009 January 28}

\begin{abstract}

VSI (`vanishing scalar invariant') spacetimes have zero values for all total
scalar contractions of all polynomials in the Riemann tensor and its covariant
derivatives.  However, there are other ways of concocting local scalar
invariants (nonpolynomial) from the Riemann tensor that need not vanish even in
VSI spacetimes, such as Cartan invariants.  Simple examples are given that
reduce to the squared amplitude for a linearized monochromatic plane
gravitational wave.  These nonpolynomial local scalar invariants are also
evaluated for non-VSI spacetimes such as Schwarzschild and Kerr and are
estimated near the surface of the earth.  Similar invariants are defined for
null fluids and for electromagnetic fields.

\end{abstract}


\maketitle

\section{Introduction}

VSI or `vanishing scalar invariant' spacetimes have recently been discussed
\cite{PPCM,C,CMPPPZ,CMPP,PCMPP,FP,CFHP,CFHP2,CGHP} and defined \cite{C} as
those in which ``all of the scalar invariants constructed from the Riemann
tensor and its covariant derivatives are zero.''  This definition has the
implicit assumption that `constructed' means by ``contraction of a
polynomial in the Riemann tensor and its covariant derivatives''
\cite{PPCM}.

However, there are other ways of forming local scalar invariants,
invariants under all passive diffeomorphisms, as defined at Eq. (1.1) of
\cite{BP}.  For example
\cite{Cartan,Ku,Brans,Karlhede,AS,Ka,Ko,O,S1,S2,SKMHH}, there are the
Cartan invariants, which are components of the Riemann tensor and its
covariant derivatives (`curvature tensors') in a frame sufficiently
determined by such curvature tensors that these components are invariant
under any remaining freedom in the frame. (One might say that these other
ways are `concoctions,' in contradistinction to the limitations of
`constructions' as implicitly defined above.)

Here I shall give simple examples from a null vector direction determined
by curvature tensors (which can be one of the null basis vector directions
of a null frame given by the Cartan method, so my examples can also be
obtained by that general method).  The first examples I shall give are
local numerical invariants $\mathcal{N}_1$ and $\mathcal{N}_2$ that are
rather simply defined for generic spacetimes (though often are either zero
or infinite). These are generically nonzero for at least certain classes of
VSI spacetimes, such as plane gravitational waves.  These examples are
concocted by taking the minimum and the maximum, over variations of an
auxiliary null vector, of the ratio of two scalar monomials each formed by
the contraction of copies of the Riemann tensor or of its covariant
derivatives and eight copies of the auxiliary null vector.  For a
linearized monochromatic plane gravitational wave, both of these local
scalar invariants are the same and are equal to the sum of the squares of
the wave amplitudes for the two polarizations, $h_+^2 + h_\times^2$.  In
this case the average value of $\mathcal{N}_1 = \mathcal{N}_2$ may also be
interpreted as eight times the number of gravitons per wavelength along the
direction of the wave and per Planck perpendicular area.

The values of these invariants are also analyzed in the Schwarzschild and Kerr
spacetimes, even though these are not VSI spacetimes and do have nonzero scalar
polynomial invariants, such as the Kretschmann invariant.  In these non-VSI
spacetimes, the nonpolynomial invariants $\mathcal{N}_1$ and $\mathcal{N}_2$
cannot be interpreted as the sum of squared of gravitational wave amplitudes or
as the number of gravitons per wavelength and per perpendicular area.

Other examples are given that can be nonzero for null fluids.  A third pair
is exhibited that are generically nonzero for plane electromagnetic waves
and give the number of photons per wavelength and per perpendicular area
for monochromatic waves.

\section{First two invariants}

For the first pair of invariants, $\mathcal{N}_1$ and $\mathcal{N}_2$,
define the covariant quantities
\begin{equation}
\nu(\mathbf{k}) \equiv R^i_{\ ajb}R^j_{\ cid} k^a k^b k^c k^d,
\label{numerator}
\end{equation}
\begin{equation}
\delta(\mathbf{k}) \equiv
R^i_{\ ajb;cd}R^j_{\ eif;gh} k^a k^b k^c k^d k^e k^f k^g k^h,
\label{denominator}
\end{equation}
which for a given event in a given spacetime are functions of an auxiliary
null vector $\mathbf{k}$ with contravariant components $k^a$.  Then define
the
$\mathbf{k}$-dependent local fraction
\begin{eqnarray}
\!\!\!\!\! f(\mathbf{k}) \!\!\! &\equiv& \!\!\!
\frac{[\nu(\mathbf{k})]^2}{2 \delta(\mathbf{k})}
\nonumber \\
&\equiv& \!\!\! \frac
{R^i_{\ ajb}R^j_{\ cid}R^m_{\ enf}R^n_{\ gmh} k^a k^b k^c k^d k^e k^f k^g k^h}
{2R^i_{\ ajb;cd}R^j_{\ eif;gh} k^a k^b k^c k^d k^e k^f k^g k^h}.
\label{frac}
\end{eqnarray}

Because there are the same number of null vector components in the
numerator as in the denominator, the ratio $f(\mathbf{k})$ is invariant
under rescaling $\mathbf{k}$, so it only depends upon the direction of
$\mathbf{k}$ over the unit sphere of directions on the null cone.  Assuming
that the denominator is generically nonzero for an arbitrary $\mathbf{k}$,
the fact that the unit sphere is compact implies that the ratio has a
minimum, which is defined to be the value of $\mathcal{N}_1$.  If (but not
only if) the denominator is nonzero for all nonzero $\mathbf{k}$ at a given
event, then the ratio also has a maximum, which is defined to be the value
of $\mathcal{N}_2$.  That is, one may define the two local scalar
invariants to be
\begin{equation}
\mathcal{N}_1 \equiv \min_\mathbf{k} f(\mathbf{k}),\ 
\mathcal{N}_2 \equiv \max_\mathbf{k} f(\mathbf{k}).
\label{def}
\end{equation}

If in $D$ dimensions one goes to a frame in which $\mathbf{k}$ is one of
two null basis vectors and the remaining $D-2$ basis vectors are
orthonormal spatial vectors orthogonal to both of the null basis vectors,
then in both Eqs. (\ref{numerator}) and (\ref{denominator}) the nonzero
terms in the Einstein summation over the indices $i$ and $j$ are restricted
to these spatial vectors, so that both $\nu(\mathbf{k})$ and
$\delta(\mathbf{k})$ are sums of perfect squares.  Since each tensor in the
product is symmetric in $i$ and $j$, there are generically $(D-2)(D-1)/2$
independent squares in each sum.  This is larger than the $D-2$ parameters
for the direction of $\mathbf{k}$ for $D\ge 4$, so generically in
spacetimes with four or more directions, both the numerator and the
denominator of the ratio $f(\mathbf{k})$ are positive for all nonzero
$\mathbf{k}$.  This allows both $\mathcal{N}_1$ and $\mathcal{N}_2$ to be
defined as finite positive local scalar curvature invariants for generic
spacetimes of at least four dimensions.

However, in a spacetime that solves the vacuum Einstein equations, with the
Ricci tensor $R_{ab}$ being either zero (the case with no cosmological constant)
or proportional to the metric (when the cosmological constant is nonzero), the
trace of each tensor in the product in either $\nu(\mathbf{k})$ or
$\delta(\mathbf{k})$ is zero when $i$ is set equal to $j$ and summed, so then
there are only $[(D-2)(D-1)/2]-1$ independent squares in each sum, which equals
the $D-2$ number of directional parameters of $\mathbf{k}$ for $D=4$. 
Therefore, in a four-dimensional vacuum spacetime, generically both the
numerator and the denominator of the ratio $f(\mathbf{k})$ can be zero for
discrete directions of $\mathbf{k}$.  Indeed, if $\mathbf{k}$ is one of the
principal null directions (PNDs) of the Weyl tensor, which always exist in four
dimensions, then the numerator will vanish \cite{Macpriv}.  Unless the
denominator also happens to vanish for the same null direction, one gets
$\mathcal{N}_1=0$, the generic case for a vacuum four-dimensional spacetime.  On
the other hand, there may or may not be null directions for which the
denominator vanishes in a vacuum four-dimensional spacetime.  Both possibilities
are generic, since the number of equations to be solved, 2, matches the number
of parameters for null directions, so that one can have generic vacuum
spacetimes, or regions of a spacetime, in which $\mathcal{N}_2$ is infinite, and
other generic vacuum spacetimes or regions in which $\mathcal{N}_2$ is finite.

In a higher-dimensional vacuum spacetime \cite{Macpriv}, the numerator will
vanish if and only if $\mathbf{k}$ is a Weyl aligned null direction
(WAND) \cite{CMPP,MCPP,PP,C2}, the generalization of a PND to higher
directions.  However, for $D>4$, only algebraically special spacetimes have
WANDs, so generically even in vacuum the numerator of $f(\mathbf{k})$ will
never vanish, and neither will the denominator.

One may also note that in a coordinate system in which the coordinates all
have the dimension of length, so that the metric components are all
dimensionless, the combinations of components of the Riemann tensor and its
covariant derivatives in both the numerator and denominator of
$f(\mathbf{k})$ have the dimensions of length to the negative eighth power,
so the ratio $f(\mathbf{k})$, and hence $\mathcal{N}_1$ and
$\mathcal{N}_2$, are dimensionless, pure numerical quantities without any
units.  Generically when they are neither zero nor infinite, their
magnitudes give two measures of the strength of the gravitational field,
increasing for stronger gravitational fields.  However, even in arbitrarily
weak gravitational fields, it is possible to have locations in spacetime
where the denominator, $2 \delta(\mathbf{k})$, of the fraction
$f(\mathbf{k})$, for all $\mathbf{k}$, is always anomalously small (or even
zero) in comparison with the numerator, $[\nu(\mathbf{k})]^2$, leading to
large (or even infinite) values for $\mathcal{N}_1$.  As noted above,
$\mathcal{N}_2$ has greater possibilities of being infinite, since it
diverges at any location where the denominator vanishes for some
$\mathbf{k}$ for which the numerator does not vanish.  Therefore,
$\mathcal{N}_1$ and $\mathcal{N}_2$ are not always good measures of the
strength of the gravitational field, as they can be arbitrarily large or
even undefined (infinite), but generically they are finite and small for
weak gravitational fields.

\section{Plane gravitational waves}

Plane gravitational waves are one class of four-dimensional vacuum
spacetimes which generically give nonzero $\mathcal{N}_1$.  In fact,
Malcolm MacCallum has suggested \cite{Macpriv} that Type N spacetimes, with
all principal null directions coinciding, may be the only such examples
(``exercise for the readers!'').  They are also examples of VSI spacetimes,
with all scalar polynomial curvature invariants vanishing, even though it
has long been known \cite{Macpriv,Ku} that they have nonzero Cartan
curvature invariant scalars.

A general plane gravitational wave that is an exact solution of the vacuum
Einstein equations has the metric \cite{BPR,EK} that can be put into the
form
\begin{equation}
ds^2 = - du dv + dx^2 + dy^2 + [(x^2-y^2)h_+''(u) + 2xyh_\times''(u)]du^2,
\label{plane}
\end{equation}
where $h_+(u)$ and $h_\times(u)$ are the wave amplitude functions for the
two polarizations, functions only of the null coordinate $u$, and where the
prime denotes a derivative with respect to that null coordinate.  Geodesics
in this metric obey the equations 
\begin{eqnarray}
x'' &=& (xh_+''+yh_\times''), \nonumber \\
y'' &=& (-yh_+''+xh_\times''). 
\label{geoeq}
\end{eqnarray}

When the wave amplitudes are small, there exists a congruence of geodesics
in which, for some period of time, the transverse coordinates $x$ and $y$
have the approximate form
\begin{eqnarray}
x(u) &\approx& x_0 + x_0 h_+(u) + y_0 h_\times(u), \nonumber \\
y(u) &\approx& y_0 - y_0 h_+(u) + x_0 h_\times(u), 
\label{geod}
\end{eqnarray}
where $x_0$ and $y_0$ are constants for each geodesic in the congruence. 
These obviously obey Eqs. (\ref{geoeq}) when the small deviations from
$(x_0,y_0)$ are dropped on the right hand sides of Eqs. (\ref{geoeq}). 
(However, including the deviations of $x$ and $y$ on the right hand sides
leads to long-term average accelerations of $x$ and $y$ toward the origin
as a result of the gravitational attraction of the energy contained in the
gravitational wave.)

The behavior of this congruence of geodesics illustrates how $h_+(u)$ and
$h_\times(u)$ are the dimensionless amplitudes for the waves.  For example,
if one has two particles in the congruence that are separated purely in the
$x$-direction, their proper separation is $\Delta x \approx \Delta x_0 (1 +
h_+(u))$, which oscillates as an oscillating wave goes by from the $h_+$
polarization.  Similarly, two particles separated purely in the $y$
direction have proper separation $\Delta y \approx \Delta y_0 (1 -
h_+(u))$, which oscillates $180^\circ$ out of phase from the same $h_+$
polarization.  On the other hand, two particles separated along a line
bisecting the $x$- and $y$-axes, say with $\Delta x_0 = \Delta y_0$, have
proper separation $(\Delta x + \Delta y)/\sqrt{2} \approx [(\Delta x_0 +
\Delta y_0)/\sqrt{2}](1 + h_\times(u))$ from the other polarization.

For the exact plane wave metric Eq. (\ref{plane}), one can readily calculate
that the nonzero components of the Riemann tensor are
\begin{eqnarray}
R_{uxux} &=& -R_{uyuy} = h_+''(u), \nonumber \\
R_{uxuy} &=& +R_{uyux} = h_\times''(u),
\label{Riem}
\end{eqnarray}
plus those related by the symmetries $R_{abcd} = R_{[ab][cd]}$. 
Furthermore, the only nonzero covariant derivatives of the Riemann tensor
are those with respect to $u$, which are the same as the partial
derivatives with respect to $u$.  As a result, one can quickly calculate
that the numerator of the ratio $f(\mathbf{k})$ in Eq. (\ref{frac}) is
$4(h_+''^2 + h_\times''^2)^2(k^u)^8$, and the denominator is $4(h_+''''^2 +
h_\times''''^2)(k^u)^8$.  Therefore, the ratio $f(\mathbf{k})$ is
independent of the auxiliary null vector $\mathbf{k}$ (though it is
strictly speaking undefined if one takes $\mathbf{k} = \partial/\partial
v$, the propagation null vector of the wave, which has $k^u = 0$, but one
can easily avoid that special direction for $\mathbf{k}$).  One immediately
gets that both local scalar numerical invariants are
\begin{equation}
\mathcal{N}_1 = \mathcal{N}_2
 = \frac{(h_+''^2 + h_\times''^2)^2}{h_+''''^2 + h_\times''''^2}.
\label{planevalue}
\end{equation}

MacCallum has informed me \cite{Macpriv} that this value is the reciprocal
of the squared modulus of a Cartan invariant calculated by Jan {\AA}man and
put into the CLASSI \cite{MS} file wave2.nul on June 9, 1986.  So this
invariant is not new, but its presentation as a simple function defined for
generic spacetimes may be.

If both $h_+(u)$ and $h_\times(u)$ are monochromatic sinusoidal functions
with the same frequency, e.g.,
\begin{eqnarray}
h_+(u) = A_+\cos{(\omega u + \phi_+)}, \nonumber \\
h_\times(u) = A_\times\cos{(\omega u + \phi_\times)},
\label{h}
\end{eqnarray}
then one simply has
\begin{equation}
\mathcal{N}_1 = \mathcal{N}_2 = h_+^2 + h_\times^2,
\label{monovalue}
\end{equation}
the sum of the two $u$-dependent oscillating squared amplitudes for the two
polarizations of the monochromatic plane wave.

For linearized plane waves, the effective average energy density of the
waves, in a local Lorentz frame in which the null coordinate of propagation
of the waves is $u = t - z$, is \cite{MTW}, in units with $G = c = 1$, the
average of $(h_+'^2 + h_\times'^2)/(16\pi)$, which for monochromatic waves
of frequency $\omega$ is the average of $\omega^2(h_+^2 + h_\times^2)/
(16\pi) = \mathcal{N}_1\omega^2/(16\pi)$.  Over one wavelength
$2\pi/\omega$ in the $z$-direction, the average energy per transverse area
(in the $x-y$ plane) is thus the average of $\mathcal{N}_1\omega/8$ over
one wavelength.  Then since each graviton of the monochromatic wave has
energy $\omega$ in units with $\hbar=1$, in Planck units ($\hbar = c = G =
1$) the number of gravitons per wavelength and per transverse area is
simply the average of $\mathcal{N}_1/8$.  That is, the value of
$\mathcal{N}_1 = \mathcal{N}_2$ averaged over one wavelength of a
monochromatic linearized plane gravitational wave is eight times the number
of gravitons per wavelength and per transverse area.

Although the number of gravitons per wavelength and per transverse area is
a Lorentz-invariant quantity in the case of a uniform flux of monochromatic
gravitons all traveling in the same direction, this number is a nonlocal
quantity (since the gravitons are not localized).  However, for classical
gravitational fields, the numerical scalar invariants $\mathcal{N}_1$ and
$\mathcal{N}_2$ defined by Eq. (\ref{def}) have the advantage of being
purely local.  They are well defined for all locations where the
denominator of $f(\mathbf{k})$ in Eq. (\ref{frac}) is nonzero for all
possible auxiliary null vectors $\mathbf{k}$, and also when the denominator
vanishes for some particular null vectors provided that the numerator also
vanishes just as fast at those null vectors (as in the plane wave example
above).

One can have situations in which the numerical scalar invariants
$\mathcal{N}_1$ and $\mathcal{N}_2$ are not defined (or are infinite), if
the denominator of the ratio $f(\mathbf{k})$ given in Eq. (\ref{frac}) is
zero for all possible auxiliary null vectors $\mathbf{k}$ (e.g., in de
Sitter) for $\mathcal{N}_1$ or goes to zero faster than the numerator does
as one approaches some special null vectors for $\mathcal{N}_2$.  (The fact
that $\mathcal{N}_1$ is defined as the minimum of the ratio means that if
the denominator, which is always nonnegative, is nonzero for any
$\mathbf{k}$, $\mathcal{N}_1$ will be finite, though $\mathcal{N}_2$, the
maximum of the ratio $f(\mathbf{k})$, would be infinite at any location
where the denominator is zero while the numerator remains nonzero for some
$\mathbf{k}$, or where both go to zero but the denominator goes to zero
faster as $\mathbf{k}$ approaches a value where the denominator vanishes.)

For the plane waves considered above, this vanishing of the denominator for
all $\mathbf{k}$ is indeed the case if one has $h_+''''^2 + h_\times''''^2
= 0$, where the 4th derivatives of both polarization amplitudes vanish.  If
one has a monochromatic wave, the numerator would also vanish at the same
values of $u$, so that one would get a finite limit of the ratio (actually
zero in the monochromatic case) if one approached it from positions where
the denominator did not vanish, but if one has a non-monochromatic wave,
one can have nongeneric situations in which $h_+''''^2 + h_\times''''^2$
vanishes but $h_+''^2 + h_\times''^2$ remains nonzero, which would make
$\mathcal{N}_1$ and $\mathcal{N}_2$ infinite.  For example, one could have
$h_+(u) = h_\times(u) = 16A \cos{(\omega u)} - A \cos{(2\omega u)}$ at
$\omega u = 2\pi n$ for any integer $n$ (hypersurfaces of measure zero in
the entire spacetime).  Or, one could have a very special spacetime in
which, say, $h_+(u) = h_\times(u) = B u^2$, which would make
$\mathcal{N}_1$ and $\mathcal{N}_2$ infinite everywhere. However, for a
generic position in a generic nonvacuum spacetime of at least four
dimensions (as well as for at least the plane wave VSI spacetimes),
$\mathcal{N}_1$ and $\mathcal{N}_2$ are well defined and are both nonzero
and finite.  (For generic vacuum spacetime in four dimensions,
$\mathcal{N}_1$ is zero, and $\mathcal{N}_2$ may be either finite or
infinite.)

\section{Schwarzschild metric}

For the Schwarzschild metric, the numerator, $[\nu(\mathbf{k})]^2$, of
$f(\mathbf{k})$ given in Eqs. (\ref{numerator}) and (\ref{frac}) is fairly
straightforward to evaluate by hand (though it did take me several hours to
calculate it, check it, and correct a mistake I made originally), but the
denominator appeared to be beyond my ability to calculate by hand in any
reasonable time.  Therefore, I used GRTensorII Version 1.79 (R4)
\cite{MPL}, which also took me several hours, with help from James
MacKinnon and Andrei Zelnikov, to install it and learn the basics of using
it, so it would not have been worth the effort just for the numerator
alone.  GRTensor enabled me to find that the numerator was exactly as
calculated by hand (0.07 seconds CPU time) and to calculate the denominator
(44.67 seconds CPU time, over 600 times as long).

Initially I did the calculation in the standard orthonormal frame outside
the event horizon, with the spatial part of the null vector making an angle
$\alpha$ with the radial direction in a local Lorentz frame of an observer
at rest with respect to the black hole.  After some simplification by hand
of the result from GRTensor, I got
\begin{equation}
f(\mathbf{k}) = \left[\frac{6M}{5r}
\frac{1}{1+\left(1-\frac{2M}{r}\right)\left(12\csc^2{\alpha}-15\right)}
\right]^2.
\label{Schwfrac}
\end{equation}

To get $f(\mathbf{k})$ both inside and outside the event horizon, it is
better to use ingoing Eddington-Finkelstein \cite{E,F,MTW} coordinates, in
which the Schwarzschild metric is
\begin{equation}
ds^2 = -\left(1-\frac{2M}{r}\right)dv^2 + 2dvdr + r^2 d\theta^2 +
r^2\sin^2{\theta}d\phi^2.
\label{Schwmetric}
\end{equation}
Then I used the one-parameter family of null vectors at each location,
\begin{equation}
\mathbf{k} = \partial_v
+ \frac{1}{2}\left(1-\frac{2M}{r}-x^2\right)\partial_r
+ \frac{x}{r}\partial_\theta,
\label{k}
\end{equation}
where $x$ is the one parameter (sufficient for a generic null vector for
Eq. (\ref{frac}), because of the spherical symmetry of the metric and
because of the independence of $f(\mathbf{k})$ on the magnitude of the null
vector $\mathbf{k}$).  The parameter $x$ may take any real value.  Outside
the horizon, $x = \sqrt{1-2M/r}\cot{(\alpha/2)}$, but the null vector
$\mathbf{k}$ given by Eq. (\ref{k}) is well defined for any real $x$ even
on and inside the horizon, i.e., for all $r>0$.

With this form of the Schwarzschild metric and of the null vector, the
hand-simplified result of GRTensor for the fraction $f(\mathbf{k})$ defined
by Eq. (\ref{frac}) is
\begin{equation}
f(\mathbf{k})\!=\!\left(\frac{2M}{5r}\right)^2\!
\left[x^2\!-\!\left(\frac{8}{3}\!-\!\frac{6M}{r}\right)\!
+\!\left(1\!-\!\frac{2M}{r}\right)^2\!\frac{1}{x^2}\right]^{\!-2}.
\label{SchwEFfrac}
\end{equation}
One can readily check that outside the horizon, using the relation given in
the previous paragraph between the parameters $x$ and $\alpha$, that
$f(\mathbf{k})$ given by Eq. (\ref{Schwfrac}) agrees with $f(\mathbf{k})$
given by Eq. (\ref{SchwEFfrac}).

The fraction $f(\mathbf{k})$ given by Eq. (\ref{SchwEFfrac}), a perfect
square for the Schwarzschild metric, goes to zero when $x$ is taken to
infinity (and also when $x$ is taken to zero at all locations except at the
event horizon, $r=2M$).  Therefore, for all locations in the Schwarzschild
metric,
\begin{equation}
\mathcal{N}_1 \equiv \min_\mathbf{k} f(\mathbf{k}) = 0.
\label{def1Schw}
\end{equation}
This vanishing of $\mathcal{N}_1$ appears to be a generic feature of
four-dimensional vacuum metrics, as we have seen above.

On the other hand, $\mathcal{N}_2$, the maximum of $f(\mathbf{k})$ over all
null vectors $\mathbf{k}$, depends on the location.  For $r\ge 3M$,
$f(\mathbf{k})=\infty$ at
\begin{equation}
x^2 = \frac{4}{3}-\frac{3M}{r}\pm
\sqrt{\left(\frac{1}{3}-\frac{M}{r}\right)
\left(\frac{7}{3}-\frac{5M}{r}\right)},
\label{x2}
\end{equation}
so for $r\ge 3M$,
\begin{equation}
\mathcal{N}_2 \equiv \max_\mathbf{k} f(\mathbf{k}) = \infty.
\label{def2Schw}
\end{equation}

For $r \le 3M$, the maximum for $f(\mathbf{k})$ is at
\begin{equation}
x^2 = \left|1-\frac{2M}{r}\right|,
\label{x3}
\end{equation}
so for $2M \le r \le 3M$,
\begin{equation}
\mathcal{N}_2 \equiv \max_\mathbf{k} f(\mathbf{k}) =
\left[\frac{3M}{5(3M-r)}\right]^2,
\label{def3Schw}
\end{equation}
whereas for $0 < r \le 2M$,
\begin{equation}
\mathcal{N}_2 \equiv \max_\mathbf{k} f(\mathbf{k}) =
\left[\frac{3M}{5(15M-7r)}\right]^2.
\label{def4Schw}
\end{equation}
At the horizon itself ($r=2M$), $\mathcal{N}_2 = 9/25 = 0.36$.

In summary, for the Schwarzschild metric, $\mathcal{N}_1$, the minimum at
each location, over all null vectors $\mathbf{k}$, of $f(\mathbf{k})$ given
by Eq. (\ref{frac}), is zero everywhere, whereas $\mathcal{N}_2$, the
maximum of $f(\mathbf{k})$, is infinite for $r \ge 3M$ but is finite for $r
< 3M$, given by Eqs. (\ref{def3Schw}) and (\ref{def4Schw}) immediately
above for $2M \le r \le 3M$ and $0 < r \le 2M$ respectively.

\section{Kerr metric}

For the Kerr metric with an orthonormal null basis
$(\mathbf{l},\mathbf{n},\mathbf{m},\bar{\mathbf{m}})$ aligned along the
eigendirections of the Weyl tensor and normalized so that
$\mathbf{l}\cdot\mathbf{n}=-1$ and $\mathbf{m}\cdot\bar{\mathbf{m}}=+1$ (and
with all other scalar products vanishing), and with the
one-complex-parameter ($z$) family of auxiliary null vectors at each
location being
\begin{equation}
\mathbf{k} 
= \mathbf{l}+z\bar{z}\mathbf{n}+z\mathbf{m}+\bar{z}\bar{\mathbf{m}},
\label{kKerr}
\end{equation}
one may calculate by hand and by GRTensor (0.13 seconds CPU time) that the
numerator of $f(\mathbf{k})$ given in Eq. (\ref{frac}) is
\begin{equation}
(R^i_{\ ajb}R^j_{\ cid} k^a k^b k^c k^d)^2 = 72^2 |\psi_2|^4 |z|^8,
\label{Kerrnumerator}
\end{equation}
where \cite{Chandra}
\begin{equation}
\psi_2 = -\frac{M}{(r-ia\cos{\theta})^3}
\label{psi}
\end{equation}
is the only nonzero Weyl curvature scalar in the Type-D Kerr metric.

My unskilled efforts with GRTensor were unsuccessful in calculating the
denominator $2 \delta(\mathbf{k})$ of $f(\mathbf{k})$, so that I gave up
after my computer had run nearly twenty-four hours without getting an
answer.  However, Nicos Pelavas \cite{Pelpriv} and Malcolm MacCallum
\cite{Macpriv} kindly calculated the denominator for me.  Pelavas' output
takes up about one page of print and is a polynomial in $z$ and $\bar{z}$
with each having powers running from 2 through 6 (25 terms), and with
coefficients functions of $M$, $a$, $r$, and $\theta$.  As I had
conjectured, the denominator is proportional to $|z|^4$ for small $|z|$,
with a coefficient (which I could not calculate) of $2^6 3^4 5^2
M^{-4/3}|\psi_2|^{10/3}$, so for small $|z|$, $f \sim
0.04|M^2\psi_2|^{2/3}|z|^4$, which has a minimum value of $\mathcal{N}_1=0$
at $z=0$.  Thus indeed $\mathcal{N}_1$ vanishes for Kerr, as it is expected
to do for a generic vacuum four-dimensional spacetime.

Malcolm MacCallum \cite{Macpriv} used spinor notation to show how the
denominator, $2 \delta(\mathbf{k})$, can be written as the absolute square
of the second covariant derivative of a Weyl curvature spinor in a frame
with $\mathbf{k}$ being one of the null basis vectors.  He explicitly
calculated this complex second derivative and expressed it as a sum of 14
terms, a reduction from the 25 terms Pelavas had given me for its absolute
square.

From MacCallum's expression, I derived the following expression for the
fraction $f(\mathbf{k})$:
\begin{equation}
f(\mathbf{k}) = \left|\frac{2M\rho}{5\Delta D}\right|^2,
\label{Kerrf}
\end{equation}
where $\rho \equiv \sqrt{r^2+u^2}$, $u\equiv a \cos{\theta}$, $\Delta
\equiv r^2 - 2Mr + a^2$, and
\begin{eqnarray}
D &=& \left(|z|-\frac{1}{|z|}\right)^2
   +A^2\left(i\sqrt{\frac{z}{\bar{z}}}-i\sqrt{\frac{\bar{z}}{z}}\right)^2
   -\frac{2}{3}\frac{P}{\rho^2\Delta} \nonumber \\
  &\ & +\frac{2}{3}i\left[A\left(\frac{3}{z}-3z
                                +\frac{4}{\bar{z}}-4\bar{z}\right)
  -\frac{Q}{\rho^2\Delta}\right] \nonumber \\
  &=& 4\cot^2{\alpha} -\frac{4}{3}A\csc{\alpha}\sin{\beta}
    +4A^2\sin^2{\beta} -\frac{2}{3}\frac{P}{\rho^2\Delta} \nonumber \\
  &\ & +\frac{2}{3}i\left[14A\cot{\alpha}\cos{\beta}
                          -\frac{Q}{\rho^2\Delta}\right],
\label{D}
\end{eqnarray}
with $z = \tan{(\alpha/2)}e^{i\beta}$, where $\alpha$ is the angle between the
spatial part of $\mathbf{k}$ and the radial direction and $\beta$ is the angle
between the nonradial spatial part of $\mathbf{k}$ and the $\theta$ direction in
the orthonormal frame determined by the orthonormal null basis,
$\mathbf{e}_{\hat{0}}=(\mathbf{l}+\mathbf{n})/\sqrt{2}$,
$\mathbf{e}_{\hat{r}}=(\mathbf{l}-\mathbf{n})/\sqrt{2}$,
$\mathbf{e}_{\hat{\theta}}=(\mathbf{m}+\bar{\mathbf{m}})/\sqrt{2}$,
$\mathbf{e}_{\hat{\phi}}=(\mathbf{m}-\bar{\mathbf{m}})/(i\sqrt{2})$, with the
null boost freedom in $\mathbf{l}$ and $\mathbf{n}$ and the rotational freedom
in $\mathbf{m}$ and $\bar{\mathbf{m}}$ used to make $\mathbf{e}_{\hat{r}}$ and
$\mathbf{e}_{\hat{\theta}}$ orthogonal to the Killing vectors
$\partial/\partial t$ and $\partial/\partial\phi$, with
\begin{equation}
A \equiv \sqrt{\frac{a^2-u^2}{\Delta}}
  \equiv \frac{a\sin{\theta}}{\sqrt{r^2-2Mr+a^2}},
\label{A}
\end{equation}
\begin{equation}
P \equiv r^4 - 3Mr^3 + 8a^2r^2 - 6u^2r^2 + Mu^2r + 8a^2u^2 - 7u^4,
\label{P}
\end{equation}
and with
\begin{equation}
Q \equiv Mu(3r^2-u^2) \equiv Ma\cos{\theta}(3r^2-a^2\cos^2{\theta}).
\label{Q}
\end{equation}

The value of $\mathcal{N}_1$, which is zero, is the minimum value of
$f(\mathbf{k})$, which corresponds to the maximum absolute value of $D$,
occurring for $|z|=0$ ($\alpha=0$) and for $|z|=\infty$ ($\alpha=\pi$). 
The value of $\mathcal{N}_2$ is the maximum value of $f(\mathbf{k})$, which
corresponds to the minimum absolute value of $D$.  For $r\gg M \ge a$ and
for $\theta$ not too near either 0 or $\pi$ (the axes of rotation), one can
always choose the complex $z$, or equivalently the direction angles
$\alpha$ and $\beta$ for the null vector $\mathbf{k}$, so that both the
real and imaginary parts of $D$ are zero.  In particular, if
$\cot{\alpha}\cos{\beta}$ is fixed to have the value $Q/(14A\rho^2\Delta)$
to make the imaginary part of $D$ vanish, so long as this value is not too
great, one can choose $\beta$ (and hence $\alpha$) to make the real part of
$D$ vanish as well.

However, if $Q/(14A\rho^2\Delta)$ is too large, then for all $\beta$ the
real part of $D$ will be positive when the imaginary part is set to zero, so
that $D$ can never be zero for any complex $z$, and $f(\mathbf{k})$ will
have a finite maximum value that is $\mathcal{N}_2$.  One can readily show
that asymptotically at large $r$ the condition for $\mathcal{N}_2$ to be
finite is $\tan{\theta} \stackrel{<}{\sim} \sqrt{27/98}(M/r)$, so that the
region with finite $\mathcal{N}_2$ is a cylinder about the axis with
asymptotic radius $\sqrt{27/98}M$.

One can also show that on the axis itself, where $A=0$, and for $r$ large
enough that $P>0$ (which in the limit of small $a$ is for $r>3M$),
\begin{equation}
\mathcal{N}_2 = \frac{9(r^2+a^2)^3}{25a^2(3r^2-a^2)^2}
=0.04\frac{r^2}{a^2}\left(1+\frac{a^2}{r^2}\right)^3
                    \left(1-\frac{a^2}{3r^2}\right)^{-2}.
\label{N2onaxis}
\end{equation}

\section{Near the surface of the earth}

If one approximated the gravitational field near the surface of the earth
by the Schwarzschild metric, then $\mathcal{N}_1$ would be zero and
$\mathcal{N}_2$ would be infinite.  However, because of perturbations of
the mass distribution from spherical symmetry, I would suspect that both
the numerator and the denominator of $f(\mathbf{k})$ given in Eq.
(\ref{frac}) would at generic nonvacuum locations be nonzero for all
directions of the auxiliary null vector $\mathbf{k}$, so that both
$\mathcal{N}_1$ and $\mathcal{N}_2$ would be nonzero and finite.  (For
vacuum locations, after the insights given by \cite{Macpriv}, I suspect that
generically $\mathcal{N}_1$ would be zero, whereas $\mathcal{N}_2$ could
generically be either finite or infinite.)

One might think that a reasonably good estimate for $f(\mathbf{k})$ and for
its minimum ($\mathcal{N}_1$) and maximum ($\mathcal{N}_2$) values at some
location would be obtained by taking the monopole plus quadrupole
contributions to the Newtonian gravitational field of the earth.  That might
indeed be correct for the numerator of $f(\mathbf{k})$ when one is far above
the surface of the earth and its higher multipole irregularities, at least
if one ignores the contribution from the gravitational fields of the sun and
moon.  However, since the denominator has each Riemann tensor with two
covariant derivatives, the small-scale variations from local mass
distributions are likely to give by far the largest contribution to the
denominator near the surface of the earth.  They are also likely to prevent
the numerator from having a minimum (say with the timelike part of the null
vector fixed) nearly so small as it would be from just the quadrupole field
of the earth.

For concreteness, let us fix the magnitude of the auxiliary null vector
$\mathbf{k}$ so that its timelike part is unity in the frame of the surface
of the earth (i.e., so that the dot product of the null vector with the
four-velocity of the earth surface is $-1$).  Then in this same frame, the
Riemann curvature tensor will have typical components of the order of
magnitude of
\begin{equation}
C = \frac{GM_\oplus}{R_\oplus^3c^2}
 \approx 1.7\times 10^{-23}\mathrm{m}^{-2},
\label{curv}
\end{equation}
where $M_\oplus$ and $R_\oplus^3c^2$ are the mass and radius of the earth. 
Nearby objects that distort the spherical gravitational field will give
distortions in the Riemann tensor of the same order of magnitude (more
nearly correctly, smaller by the ratios of the densities of the objects to
the average density of the earth, but since these ratios are of the order of
unity, here I shall ignore them).

Therefore, the magnitude of the numerator of $f(\mathbf{k})$ can be
expected to be of the order of $C^4$ for generic directions of the
spacelike part of $\mathbf{k}$.  From the results above for the
Schwarzschild metric, it would vary as $\sin^8{\alpha}$, where $\alpha$ is
the angle between the spatial part of the null vector and the radial
direction, if the gravitational field were precisely spherical and vacuum,
but the distortions of nearby objects, e.g., in one's office, would
presumably make the numerator have a magnitude of the order of $C^4$ for
generic directions of the spacelike part of $\mathbf{k}$.  However, in a
generic vacuum spacetime, the numerator would be expected to vanish as
$\sin^4{\alpha}$, where $\alpha$ is now the angle between the spatial part
of the null vector and the spatial part of the nearest principal null
direction.  In a non-vacuum region, the minimum value of the numerator
(with the magnitude of $\mathbf{k}$ fixed as above) would be expected to be
of the order of the tracefree part of the Ricci tensor, $G \rho/c^2 \sim
C\rho/\rho_\oplus$, where $\rho$ is the mass density of the location and
$\rho_\oplus$ is the average mass density of the earth.

Similarly, the magnitude of the denominator of $f(\mathbf{k})$ can be expected
to be of the order of $C^2/L^4$ for generic directions of the spacelike part of
$\mathbf{k}$, where $L$ is a typical length scale for the smallest nearby object
(at a distance comparable to its size) that is distorting the Riemann tensor of
the earth's gravitational field.  Therefore, one would expect that
$f(\mathbf{k})$ would for generic directions of the spacelike part of
$\mathbf{k}$ have magnitudes of the order of $C^2L^4$, with its maximum value
being of this same order (if it is finite) and its minimum value being of the
order of $C^2L^4\rho/\rho_\oplus$.  If one takes $L\sim 1$ m, say for the old
CRT computer monitor some fraction of a meter in front of me that I had when I
started writing this paper, and a location within the air where
$\rho/\rho_\oplus \sim 10^{-3}$, then one would expect that, very crudely,
\begin{equation}
\mathcal{N}_1 \sim 10^{-26},\ \mathcal{N}_2 \sim 10^{-23}
\label{curvvalue}
\end{equation}
(unless $\mathcal{N}_2 = \infty$) at a typical location in the air, within
a meter or so of lumpy stuff of roughly the density of the earth.  These
order-of-magnitude estimates exhibit the fact that for generic weak
nonvacuum gravitational fields, I would expect both $\mathcal{N}_1$ and
$\mathcal{N}_2$ to be very small, increasing quadratically with the
strength of the field (e.g., quadratically with the mass densities and
hence also with $C$ for a given spatial distribution of the relative
densities).

\section{Other local invariants nonzero for VSI spacetimes}

The examples of $\mathcal{N}_1$ and $\mathcal{N}_2$ defined by Eq.
(\ref{def}) as the minimum and maximum, respectively, of the ratio
$f(\mathbf{k})$ given in Eq. (\ref{frac}), are of course only two examples
of an obviously infinite sequence of such examples of local invariants that
might be nonzero for VSI spacetimes.  (Those examples were merely the
simplest examples that I could readily think of.)  All one needs is some
contraction of Riemann tensors and their covariant derivatives with null
vectors in both the numerator and denominator, so that there are the same
number of copies of each null vector in both (in order that the ratio not
depend on the normalization of each null vector, but only on its direction,
which forms a compact set allowing one to take the minimum and the maximum
of the ratio as local invariants).  One can even have arbitrarily many
different null vectors in both the numerator and denominator, so long as
each null vector occurs the same number of times in both.  Alternatively,
one can have different numbers of null vectors in the numerator and
denominator if one normalizes them so that, say, either the numerator or
the denominator is fixed to be unity.

Instead of equivalence classes of null vectors related by scalar
multiplication for the objects to contract with the Riemann tensors and
their covariant derivatives, one could alternatively take equivalence
classes of other geometric objects that are equally contracted with Riemann
tensor objects in both the numerator and the denominator, at least so long
as each equivalence class forms a compact set so that one can define the
maximum and minimum of the resulting ratio.

Of course, it might be difficult to evaluate more complex examples.  Even
the simple examples given here appear rather intractable much beyond
the simple cases considered above of a plane gravitational wave and of the
Schwarzschild metric.  However, their conceptual possibility shows that
there are far more local invariants for a gravitational field that may be
concocted than just the polynomial invariants that are normally
constructed.

However, all of these local scalar invariants are just special cases of
what can be constructed from the Cartan invariants, since they are
determined by components of the curvature and of its covariant derivatives
in special bases either partially or totally determined by the curvature
and its derivatives.  Therefore, they are not logically independent methods
of concocting other invariants beyond those that may be obtained from the
Cartan invariants, but some of them might be thought of as somewhat simpler
concoctions that shortcut the most general form of the Cartan procedure.

It might also be worthwhile being reminded that if one does not restrict
attention to local invariants, as was done here, then there would be an even
vastly greater set of invariants.  For example, even for flat spacetime, if
space is compactified into a torus, then the periods and angles of the torus
would be nonlocal invariants.  When there is curvature, one can readily
think of even many more ways to construct nonlocal invariants.

\section{Invariants for null fluids and electromagnetic fields}

The examples of $\mathcal{N}_1$ and $\mathcal{N}_2$ defined above by Eq.
(\ref{def}) were concocted to be nonzero for generic spacetimes, and also
for generic plane waves that are VSI spacetimes, though apparently
$\mathcal{N}_1$ is zero for generic four-dimensional vacuum spacetimes. 
However, one can have VSI spacetimes that are not vacuum but have null
fluids, electromagnetic waves, dilatons, and/or supergravity fluxes
\cite{PPCM,C,CMPPPZ,CMPP,FP,CFHP,CFHP2,CGHP}, for which in generic
circumstances one can concoct nonzero local invariants that are simpler
than $\mathcal{N}_1$ and $\mathcal{N}_2$.

For example, instead of the ratio $f(\mathbf{k})$ defined by Eq.
(\ref{frac}), for a nonvacuum spacetime one could define
\begin{equation}
\hat{f}(\mathbf{k}) \equiv \frac
{R_{ab}R_{cd} k^a k^b k^c k^d}
{R_{ab;cd} k^a k^b k^c k^d}
\label{frac2}
\end{equation}
in terms of the Ricci tensor $R_{ab}$ (or in terms of its tracefree part,
$S_{ab} = R_{ab}-\frac{1}{D}Rg_{ab}$ in $D$ dimensions, which would give
the same result) and then take
\begin{equation}
\mathcal{N}_3 \equiv \min_\mathbf{k} \hat{f}(\mathbf{k}),\ 
\mathcal{N}_4 \equiv \max_\mathbf{k} \hat{f}(\mathbf{k}).
\label{def2}
\end{equation}
Assuming that the covariant derivatives of the Ricci tensor are suitably
nonzero, these invariants may be well defined and nonzero even for generic
null fluid VSI spacetimes.

For an electromagnetic field tensor $F_{ab}$, one might define
\begin{equation}
\tilde{f}(\mathbf{k}) \equiv \frac
{F^i_{\ a} F_{ib} F^j_{\ c} F_{jd} F^k_{\ e} F_{kf} k^a k^b k^c k^d k^e k^f}
{F^i_{\ a;bc} F_{id;ef} k^a k^b k^c k^d k^e k^f}
\label{frac3}
\end{equation}
and then take
\begin{equation}
\mathcal{N}_5 \equiv \min_\mathbf{k} \tilde{f}(\mathbf{k}),\ 
\mathcal{N}_6 \equiv \max_\mathbf{k} \tilde{f}(\mathbf{k}).
\label{def3}
\end{equation}
For simplicity (though more general examples in curved spacetime would also be
interesting to investigate), let us consider the case in which the
electromagnetic field stress-energy tensor has a negligible effect upon the
spacetime, which has the nearly flat metric
\begin{equation}
ds^2 = - du dv + dx^2 + dy^2
\label{flat}
\end{equation}
with $u=t-z, v=t+z$.  Then take the example of a plane electromagnetic wave
traveling in the $z$-direction, with electromagnetic potential one-form
\begin{equation}
A = A_x(u)dx + A_y(u) dy
\label{EMA}
\end{equation}
and electromagnetic field two-form
 \begin{equation}
F = dA = A'_x(u) du\wedge dx + A'_y(u) du\wedge dy.
\label{EMF}
\end{equation}

With $u = t-z$, this gives $E_x = B_y = -A'_x$ and $E_y = - B_x = -A'_y$. 
This plane-wave electromagnetic field then gives $\tilde{f}(\mathbf{k})$
that is independent of $\mathbf{k}$ and has the form
\begin{equation}
\mathcal{N}_5 = \mathcal{N}_6
 = \frac{(A'^2_x + A'^2_y)^3}{A'''^2_x + A'''^2_y}.
\label{EMvalue}
\end{equation}

If both $A_x(u)$ and $A_y(u)$ are monochromatic sinusoidal functions
with the same frequency, e.g.,
\begin{eqnarray}
A_x(u) = a_x\cos{(\omega u + \phi_x)}, \nonumber \\
A_y(u) = a_y\cos{(\omega u + \phi_y)},
\label{Vecpot}
\end{eqnarray}
then one has
\begin{equation}
\mathcal{N}_5 = \mathcal{N}_6 = (E_x^2+E_y^2)^2/\omega^4.
\label{monoEMvalue}
\end{equation}

Since the energy density of the electromagnetic wave is
$(E_x^2+E_y^2)/(4\pi)$, over one wavelength $2\pi/\omega$ in the
$z$-direction the average energy per transverse area (in the $x-y$ plane)
is thus the average of $\sqrt{\mathcal{N}_5}\omega/2$ over one wavelength. 
Then since each photon of the monochromatic wave has energy $\omega$, the
number of photons per wavelength and per transverse area is simply the
average of $\sqrt{\mathcal{N}_5}/2$.  That is, the value of
$\sqrt{\mathcal{N}_5} = \sqrt{\mathcal{N}_6}$ averaged over one wavelength
of a monochromatic linearized plane electromagnetic wave is twice the
number of photons per wavelength and per transverse area.

\section{Conclusions}

In this paper I have given simple illustrations of the long-known fact that VSI
spacetimes (which have all scalar invariants vanishing everywhere that are
constructed as polynomials in total contractions of the Riemann tensor and its
covariant derivatives) have other classes of local scalar invariants that
generically do not vanish.  The examples in this paper are concocted by taking
the ratio of total contractions of the Riemann tensor and its covariant
derivatives with the same number of copies of an auxiliary null vector in both
the numerator and denominator and then taking the minimum or the maximum of the
ratio as the direction of the null vector is varied over its compact unit sphere
on the null cone.  When these minima or maxima are finite, they are well-defined
local Lorentz invariants of the gravitational field.  Examples were given that
reduce to the squared amplitudes of monochromatic plane gravitational waves (and
whose average over one wavelength is proportional to the number of gravitons per
wavelength and per cross-sectional area).  The first of these examples seems to
vanish for generic vacuum spacetimes in four dimensions.  Both of these example
local scalar invariants were fully calculated for the Schwarzschild metric,
partially calculated for the Kerr metric, and estimated near the surface of the
earth.  Other examples were given for null fluids or null electromagnetic
fields.

These examples illustrate the point made by the Cartan constructions
\cite{Cartan,Brans,Karlhede,AS,Ko,O,S1,S2,SKMHH} that local scalar
invariants may be concocted that are nonzero for a larger class of
spacetimes than one might na\"{\i}vely think, such as the VSI spacetimes
for which all scalar polynomial invariants in the curvature and its
derivatives vanish.

\section*{Acknowledgments}

I am grateful to Peter Musgrave, Denis Pollney and Kayll Lake \cite{MPL} for
providing GRTensor free of charge for calculating the first two invariants above
for the Schwarzschild metric, to James MacKinnon and Andrei Zelnikov for helping
me install GRTensor and use it, to Jan {\AA}man, Alan Coley, Stanley Deser,
Maciej Dunajski, Gary Gibbons, Sigbj{\o}rn Hervik, Malcolm MacCallum, Robert
Milson, Georgios Papadopoulos, Nicos Pelavas, and Vojtech Pravda for extensive
email discussions after the first version of this paper, and to two anonymous
referees for suggesting corrections and clarifications.  This research was
supported in part by the Natural Sciences and Engineering Research Council of
Canada.

\end{document}